\documentclass[aps,prl,twocolumn]{revtex4}
\usepackage{amsmath,amssymb,graphicx,bbold}
\usepackage{graphicx}  
\usepackage{dcolumn}   
\usepackage{bm}        
\usepackage{color}

\providecommand{\abs}[1]{\left\lvert#1\right\rvert}

\begin{document}

\title{Pattern revivals from fractional Gouy phases in structured light}
\author{B. Pinheiro da Silva$^1$, V. A. Pinillos$^2$, D. S. Tasca$^1$, L. E. Oxman$^1$ and A. Z. Khoury$^1$}

\affiliation{1- Instituto de F\'\i sica, Universidade Federal Fluminense,
24210-346 Niter\'oi, RJ, Brasil}
\affiliation{2- Departamento de Ciencias, Pontificia Universidad Cat\'olica del Per\'u, 
Apartado 1761, Lima, Per\'u}
\date{\today}

\begin{abstract}
We investigate pattern revivals in specially designed optical structures that combine different transverse modes. 
In general, the resulting pattern is not preserved under free propagation 
and gets transformed due to non synchronized Gouy phases. However, it is possible 
to build structures in which the Gouy phases synchronize at specific fractional values, 
thus recovering the initial pattern at the corresponding longitudinal positions. This effect is illustrated with a 
radially structured light spot in which the beam energy can be addressed to different positions without the need of 
intermediate optical components, what can be useful for optical communications and optical tweezing with structured beams.
\end{abstract}

\maketitle


The transverse structure of optical beams has been widely used for encoding information both in the classical and quantum 
domains \cite{physrep,25years}. Transmission and amplification of images and patterns either in free space or in optical resonators is still an 
active research area. In multi mode regime, the spatial structure of nonlocal quantum correlations gives rise to 
interesting and counter-intuitive effects such as the so called \textit{ghost images} \cite{ufmg,ghost,moreau}. 
When few spatial modes are involved, they can serve as a finite dimension Hilbert space for encoding quantum information \cite{sebastiao,sebastiao2,fracuffufmg}.

Within the limits of the paraxial approximation, an arbitrary spatial structure can be expanded 
in the well known Hermite-Gaussian (HG) or Laguerre-Gaussian (LG) bases that are orthonormal solutions of the paraxial wave 
equation. Apart from a transverse rescaling due to diffraction, the intensity patterns associated to these solutions 
preserve their shape along free space propagation. However, their phase evolution depends on the mode parameters 
so that intensity patterns composed by different paraxial modes may not preserve their shape in free space propagation. 
In this regard, the Gouy phase plays a crucial role affecting the relative phase among the modes taking part in 
the pattern superposition. For example, this can lead to self-rotating patterns when different orbital angular momenta 
are combined in the mode superposition, as studied in Ref.\cite{baumann}.

Free propagation of spatially modulated beams leads to surprising features such as the well known Talbot effect first 
noticed by Henry Fox Talbot in the nineteenth century \cite{talbot}. He noticed that the spatial modulation imposed 
by a diffraction grating would periodically reproduce itself in the near field pattern. This effect has been 
also realized in matter waves \cite{talbot-matter}. 
Non-diffracting and self-imaging beams have been investigated for a long period due to their potential applications 
in microscopy, micro-fabrication and optical tweezing \cite{patorsky,bottle}. These beams can be engineered by suitably shaping the transverse 
Fourier spectrum of the light field, allowing the preparation of non-diffracting Bessel beams and projection of a given 
image to a desired position in space \cite{courtial-self-image}.
In this work we show that pattern revivals can also result from a suitable superposition of discrete 
paraxial modes. In such structures the Gouy phases of the component modes synchronize at specific longitudinal positions, 
recovering the initial mode superposition apart from an irrelevant overall phase factor. The occurrence of these revivals 
is related to fractional Gouy phases determined by the mode orders. As an illustration, we exploit radially structured  
Laguerre-Gaussian modes to construct a self-restoring spot that can be used as the basic unit of a more complex pattern and 
transmit information to desired positions. 
This can be useful in different contexts, including optical imaging, tweezing and quantum information platforms.



Free space propagation of optical beams in paraxial regime can be described by two-dimensional 
Hermite-Gaussian (HG) or Laguerre-Gaussian (LG) functions. The HG basis for a beam with wave 
number $k$ propagating along the $z$ axis reads 
\begin{eqnarray}
u_{mn}(x,y,z) = X_m(x,z)\,Y_n(y,z)\,e^{-i\phi_{N}(z)}\;,
\end{eqnarray}
where the Gouy phase 
\begin{eqnarray}
\phi_{N}(z) = \left( m + n + 1\right)\,\arctan\left( z/z_0\right)\;,
\end{eqnarray}
depends on the mode order $N=m+n$ and the Rayleigh parameter $z_0$ characterizing the diffraction distance. 
The mode functions are of the form  
\begin{eqnarray}
F_m(\xi,z) &=& \frac{\mathcal{N}_{m}}{\sqrt{w}}\,e^{-\left[\frac{ik\xi^2}{2R}+\frac{\xi^2}{w^2}\right]}
\, H_m\left(\frac{\xi\sqrt{2}}{w}\right)\,,
\end{eqnarray}
where $F=X,Y\,$, $\xi=x,y$ and $H_m(\xi)$ is the Hermite polynomial of order $m\,$.
The normalization constant $\mathcal{N}_{m}\,$, the wavefront radius $R(z)$ and 
the beam width $w(z)$ are given by 
\begin{eqnarray}
\mathcal{N}_{m} &=& \frac{(2/\pi)^{1/4}}{2^{\,m/2}\,m\,!}\;,
\\
R(z) &=& \frac{z_0^2 + z^2}{z}\;,
\\
w(z) &=& \left[\frac{2\,(z_0^2 + z^2)}{k\,z_0}\right]^{1/2}\;.
\end{eqnarray}
The origin of the $z$ axis is placed on the focal plane where the beam width is minimal 
and the wavefront is plane ($R\to\infty$).

In cylindrical coordinates, paraxial modes can be expanded in the LG basis 
\begin{eqnarray}
v_{p\,l}(r,\theta,z) = R_{p\,l}(r,z)\,e^{i\,l\theta}\,e^{-i\phi_{N}(z)}\;,
\end{eqnarray}
where the Gouy phase now reads 
\begin{eqnarray}
\phi_{N}(z) = \left( 2p + \abs{l} + 1\right)\,\arctan\left( z/z_0\right)\;,
\end{eqnarray}
and the mode order is $N=2p+\abs{l}\,$. The radial function is given by  
\begin{eqnarray}
R_{p\,l}(r,z) &=& \frac{\mathcal{N}_{p\,l}}{w}\,\left(\frac{r\sqrt{2}}{w}\right)^{\abs{l}} 
L_p^{\abs{l}}\left(\frac{2r^2}{w^2}\right) e^{-\left[\frac{ik\,r^2}{2R}+\frac{r^2}{w^2}\right]}\,,
\nonumber\\
\end{eqnarray}
where $L_p^{\abs{l}}$ are generalized Laguerre polinomials and the normalization 
constant is 
\begin{eqnarray}
\mathcal{N}_{p\,l} &=& \left[\frac{2p\,!}{\pi\,(p+\abs{l})\,!}\,\right]^{1/2}\;.
\end{eqnarray}

Let us consider a normalized pattern composed by $d$ basic structures with different orders 
$N_1,N_2,\cdots,N_d\,$, 
\begin{eqnarray}
\psi (\mathbf{r}) = \sum_{j=1}^{d} A_{j}\,\varphi_{j}(\mathbf{r})\,e^{-i\phi_{N_j}(z)}\;. 
\end{eqnarray}
Each basic structure $\varphi_{j}(\mathbf{r})$ can be a combination of HG, LG or both kinds 
of modes sharing the same order $N_j\,$. Apart from diffraction, each basic structure evolves 
with a stable pattern and acquires a Gouy phase $\phi_{N_j}(z)\,$. Due to the different mode 
orders present in the superposition, the Gouy phases will evolve 
at different rates and $\psi$ will not keep the original pattern along propagation. However, 
the pattern prepared at $z=0$ may be repeated if all Gouy phases re-synchronize along propagation, 
that is, if 
\begin{eqnarray}
\phi_{N_{j+1}}(z) - \phi_{N_{j}}(z) = 2\,q_j\,\pi\;\;(q_j\in\mathbb{Z})\;.
\label{2qpi} 
\end{eqnarray}
For a non astigmatic beam in free space, this will occur at specific positions, but due to 
the limited range of the Gouy phase, not all multiples can be achieved. 

In order to derive the exact location of the revivals, we first write 
\begin{eqnarray}
\psi_d (\mathbf{r}) = e^{-i\bar{\phi}(z)}\,\sum_{j=1}^{d} A_{j}\,
\varphi_{j}(\mathbf{r})\,e^{-i\,\Delta_j(z)}\;, 
\end{eqnarray}
where 
\begin{eqnarray}
\bar{\phi}(z) &=& \frac{1}{d} \sum_{j=1}^{d} \phi_{N_j}(z)\;,
\label{phibar}
\\
\Delta_j(z) &=& \phi_{N_j}(z) - \bar{\phi}(z)\;.  
\label{Delta}
\end{eqnarray}
Note that $\sum_j \Delta_j = 0\,$, so that the phasors $e^{-i\,\Delta_j}$ form 
a diagonal $SU(d)$ matrix evolving along the beam propagation. A revival will occur 
when these phasors get aligned and this matrix structure becomes proportional to the identity, 
which can only happen with fractional phase factors since the evolution is closed in 
$SU(d)\,$\cite{fracuff}. To see this more explicitly we can use Eqs.(\ref{2qpi}) and 
(\ref{Delta}) to arrive at the fractional phase condition
\begin{eqnarray}
\sum_{j=1}^{d} \Delta_j &=& d\,\Delta_1 + 2\pi\,\sum_{j=2}^{d}\sum_{k=1}^{j-1} q_k=0\;,
\nonumber\\
\Rightarrow \Delta_j &=& \frac{2\pi r_j}{d}\qquad (r_j\in\mathbb{Z})\;,
\label{eq:fracphase}
\end{eqnarray}
where $r_1 = -\sum\limits_{k=1}^{d-1} (d-k)\,q_k$ and $r_{j+1} = r_{j} + 2\,d\,q_j\,$. 
Therefore, revivals can only be obtained when the fractional phase condition (\ref{eq:fracphase}) 
is fulfilled. 

Going back to Eq.(\ref{2qpi}), we obtain the revival positions from 
\begin{eqnarray}
\left(N_{j+1} - N_{j}\right)\,s(z) = 4\,q_j\;,  
\label{conds}
\end{eqnarray}
where $s(z)\equiv (2/\pi)\,\arctan(z/z_0)\,$. Note that as one moves from the focal plane ($z=0$) to the far 
field region ($z\to\infty$) the parameter $s$ varies continuously between 0 and 1. In practice the far field 
configuration can be approximately observed at distances $z\gg z_0\,$, that corresponds to $s\sim 1\,$.

Revivals 
require a minimum value ($N_{j+1} - N_{j} \geq 4$) for non-vanishing gaps between consecutive mode orders, 
except for the trivial situation in which $N_{j+1} = N_j\,$, corresponding to a pattern 
composed by modes of the same order that remain in phase along propagation. 
In the general case, for a given set 
of modes present in the superposition, the first revivals always occurs at the position 
where $4/s(z)$ reaches the maximum common denominator $D(N_2-N_1,\,\dots\,,N_d-N_{d-1})$ of 
\textit{all nonvanishing mode gaps}. Then, other revivals occur at the positions corresponding 
to the sub-multiples of $D\,$, reachable within the limits imposed on $s\,$. 
If the mode gaps are incommensurable, then revivals never occur, since it would 
imply $s=4\,$. 
At the revival positions, the field distribution becomes 
\begin{eqnarray}
\psi_d\left(\bar{x},\bar{y},z\right) = e^{i\,\bar{\phi}(z)}\, 
e^{\frac{2i\pi r}{d}}\, 
\psi_d\left(\bar{x},\bar{y},0\right)\;,  
\end{eqnarray}
where $\bar{x}=x/w(z)\,$ $\bar{y}=y/w(z)$ and $r=\min_j\abs{r_j}\,$.

We next present some interesting numerical examples and the corresponding experimental results 
supporting our theoretical predictions for the revivals on the fractional Gouy 
phase positions.

\begin{figure}[h!]
	\includegraphics[scale=0.31]{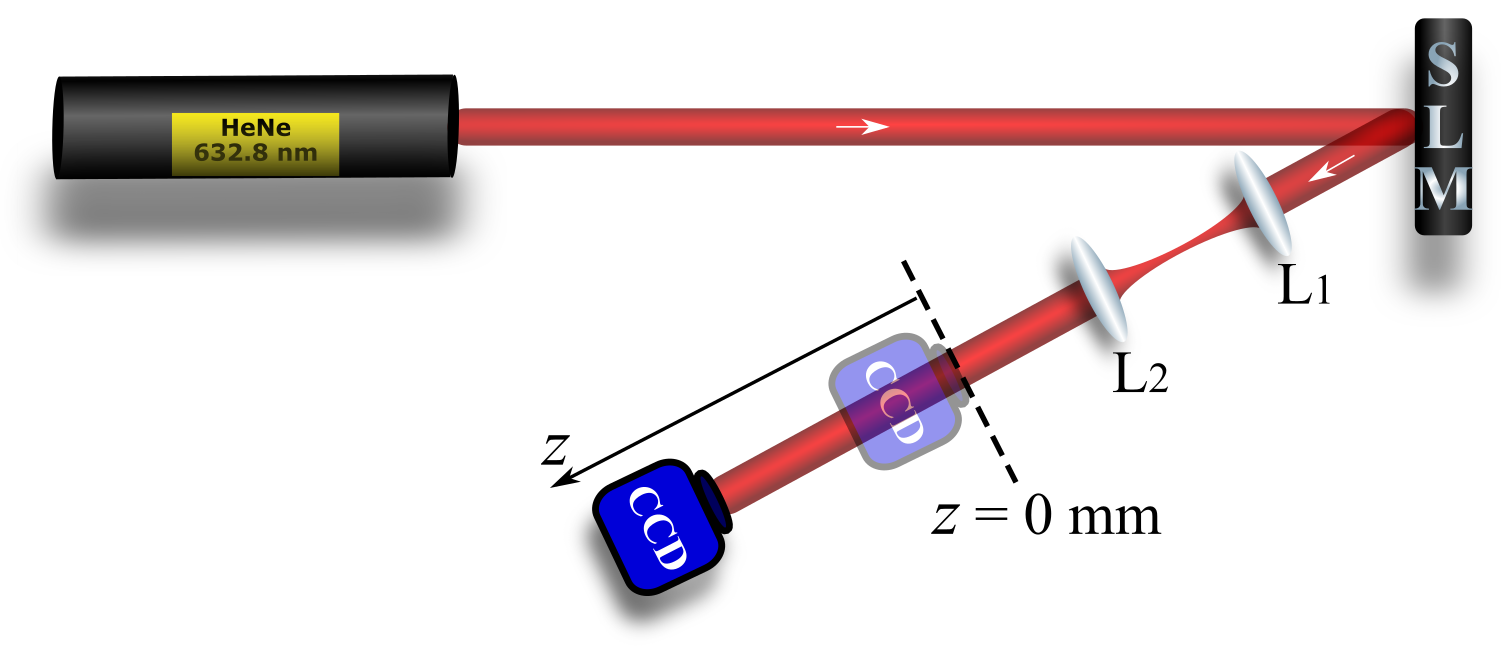}
	\caption{Experimental setup.}
	\label{fig:setup}
\end{figure}

We demonstrated the revival effect with the setup shown in Fig.\ref{fig:setup}. A Gaussian beam from a He-Ne 
laser is sent to a spatial light modulator (SLM) programmed to produce the desired mode superposition. The screen 
of the SLM is imaged to a remote position with a pair of lenses (L1 and L2) to allow acquisition of the initial 
image with a
charge-coupled device (CCD) camera. Then, the CCD is translated along the propagation axis $z\,$, starting from the 
screen image position $z=0\,$.

\begin{figure}
	\includegraphics[scale=0.16]{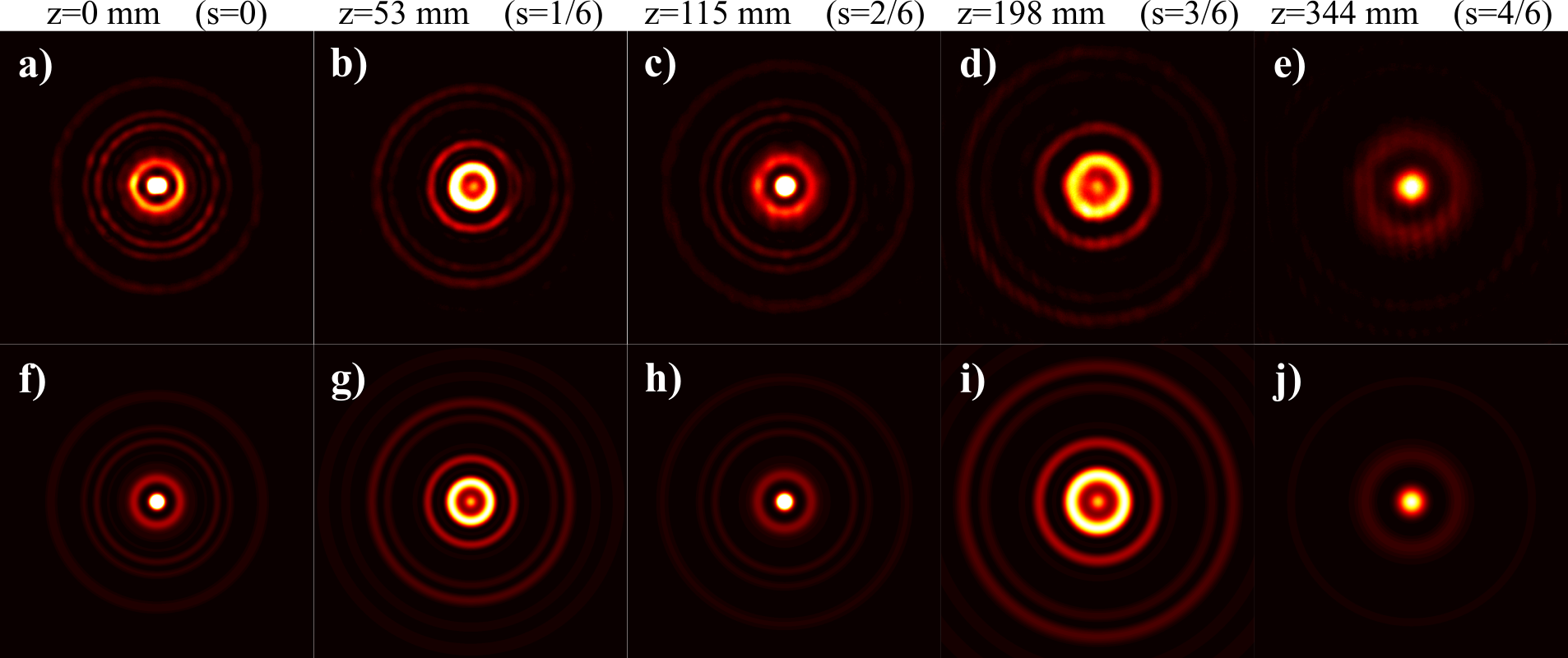}
	\caption{Measured (a)-(e) and calculated (f)-(j) intensity distribution of the structure 
		given by Eq.(\ref{eq:psi3}) at different planes along propagation.}
	\label{fig:spot}
\end{figure}

\begin{figure}[h!]
	\includegraphics[scale=0.29]{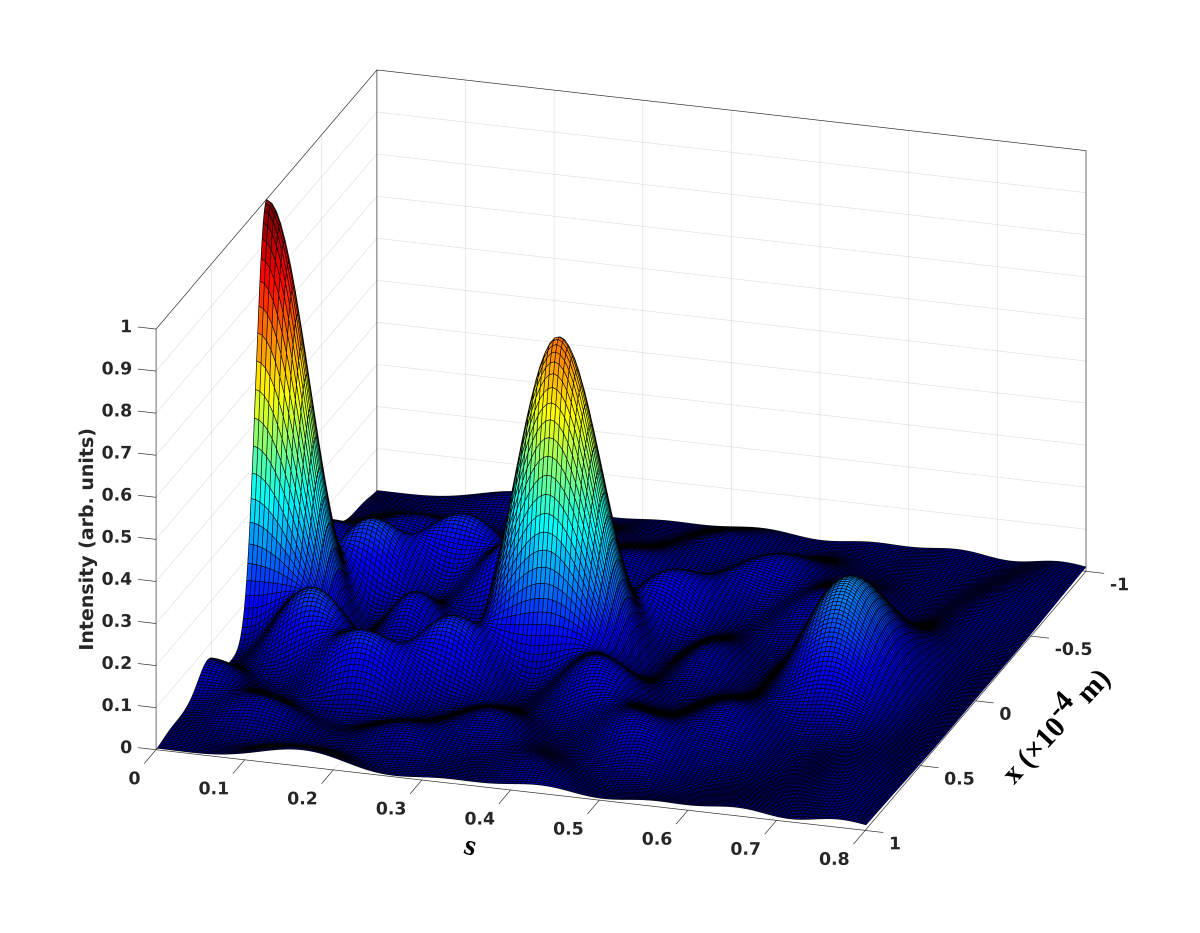}
	\caption{3D graph of the structured spot evolution, showing the spontaneous reconstitution of the intensity concentration.}
	\label{fig:spot3D}
\end{figure}

Many different examples can be produced illustrating the revival effect based on fractional Gouy phases.
We will focus our demonstration on a superposition of three radial LG modes that define a bright spot at the 
focal plane and distributes its energy along propagation. The bright spot gets restored at the fractional Gouy 
phase positions. The effect is illustrated with the following three-mode structure
\begin{eqnarray}
\psi_{3}(\mathbf{r}) = 0.3\,v_{0,0}(\mathbf{r}) + v_{6,0}(\mathbf{r}) + v_{12,0}(\mathbf{r})\;,
\nonumber\\
\label{eq:psi3}
\end{eqnarray}
carrying $l=0$ and $p=0,6,12\,$. The mode gaps involved in the structure given in (\ref{eq:psi3}) are $N_2 - N_1 = N_3 - N_2 = 12\,$, 
whose denominators are $D=1,2,3,4,6,12\,$. The revival positions are determined by $s(z_j) = 4/D_j$
within the interval $0\leq s\leq 1\,$. This results in $s(z_1) = 1/3\,$, $s(z_2) = 2/3$ and $s(z_3) = 1\,$, the last one 
corresponding  to $z_3\to\infty\,$.
We shall not care about normalization issues, since it is completely irrelevant 
to our analysis. The relative weights of the radial modes were optimized to concentrate the intensity in the central 
spot, resulting in the distribution shown in Fig.\ref{fig:spot}a where an intense central spot is surrounded by weak 
rings. As the beam propagates, this structure evolves and the intensity gets transversely redistributed as shown by 
the experimental results given in Figs.\ref{fig:spot}a-e and the corresponding theoretical results given by the density 
plots in Figs.\ref{fig:spot}f-j. The agreement between experimental and theoretical results is remarkable. 
We have also included supplemental material 1 with an animation of the structured spot evolution and the geometric 
representation of the Gouy phase variation, showing synchronization at the revival positions.
In Fig.\ref{fig:spot3D} we plot the calculated intensity distribution along a transverse diameter as a function of the 
propagation parameter $s$ in a range covering two revival positions. 
It shows the relative intensity between the central spot and the surrounding rings.

\begin{figure}[h!]
	\includegraphics[scale=0.31]{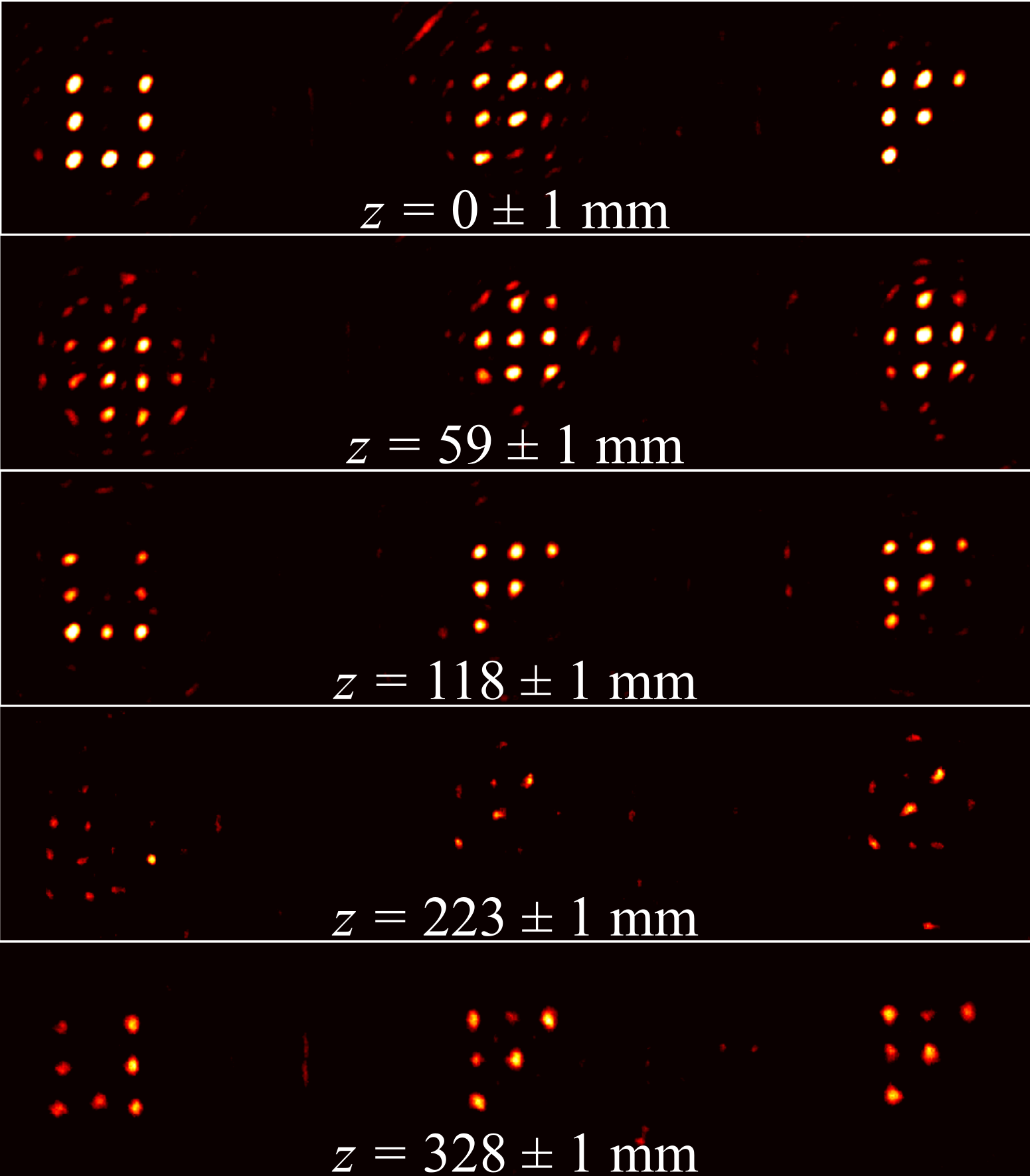}
	\caption{Propagation sequence of the structured spot pattern forming the initials UFF.}
	\label{fig:UFF}
\end{figure}

As a striking illustration of the revival effect, 
we used the SLM to implement a set of structured spots given by Eq. (\ref{eq:psi3}), centered at different locations 
disposed to form the initials UFF. The initial pattern is displayed in Fig. \ref{fig:UFF}a. Then, the patterns appearing 
at different propagation points are shown in Figs. \ref{fig:UFF}b-e. In order to improve the pattern visibility, we have 
applied a uniform step filter that cuts out the illumination caused by the weak external rings and the background light.
Two revivals are evident at the positions $z=118 mm$ and $328 mm\,$, where the Gouy phases get synchronized at the 
fractional phase differences. 
We have also filmed the pattern evolution along propagation from the initial position until the first revival. The experimental 
videos are provided in supplemental materials 2 (raw images) and 3 (noise filtered images).



In conclusion, we have demonstrated a method for structuring light with pattern revivals using 
Gouy phase synchronization at fractional values. The propagation positions of the revivals are 
determined by the integer relation between the Gouy phase and the paraxial mode order, where 
comensurability between the different modes plays an essential role. This can motivate the 
quest for more involved connections between number theory and paraxial optics \cite{gauss-sum}. 
The ability to shape the 
transverse intensity distribution and address the beam energy to specific points without the 
need of intermediate optical components can be useful for free space communication and optical tweezing. 
Moreover, the Gouy phase plays also an important role in matter waves, where the ideas presented here 
can find interesting applications in electron beam microscopy and Bose-Einstein condensates.

\section*{Acknowledgments}
Funding was provided by Conselho Nacional de Desenvolvimento Tecnol\'ogico (CNPq), 
Coordena\c c\~{a}o de Aperfei\c coamento de Pessoal de N\'\i vel Superior (CAPES), 
Funda\c c\~{a}o de Amparo \`{a} Pesquisa do Estado do Rio de Janeiro (FAPERJ), and 
Instituto Nacional de Ci\^encia e Tecnologia de Informa\c c\~ao Qu\^antica (INCT-CNPq).

\end{document}